\documentclass[12pt]{article}
\usepackage{pdproc} 

\usepackage[centertags]{amsmath}
\allowdisplaybreaks[1]
\usepackage{amsbsy}
\usepackage{amsfonts}
\usepackage{amssymb}
\usepackage[dvips]{graphicx}
\usepackage{cite}

\newcommand{\boss}[2]{\ensuremath{\rlap{\kern-2.5pt\ensuremath{\overset{\scriptscriptstyle(-)}{\phantom{#1}}}}{\ensuremath{{#1}_{#2}}}}}

\begin{document}

\renewcommand{\thefootnote}{\alph{footnote}}

\title{STERILE NEUTRINO FITS}

\author{CARLO GIUNTI}

\address{INFN, Sezione di Torino, Via P. Giuria 1, I--10125 Torino, Italy}

\abstract{
After a brief review of the results of solar, atmospheric and long-baseline neutrino oscillation experiments
which led to the current three-neutrino mixing paradigm,
we discuss indications of neutrino oscillation experiments
in favor of short-baseline oscillations
which require the existence of one or more sterile neutrinos.
We show that the simplest possibility of existence of one sterile neutrino
is not enough to fit all data of short-baseline neutrino oscillation experiments
because of two tensions:
a tension between neutrino and antineutrino data
and a tension between
appearance and disappearance data.
The tension between neutrino and antineutrino data is eliminated with the addition
of a second sterile neutrino which allows CP-violating effects
in short-baseline experiments.
In this case the tension between
appearance and disappearance data is reduced,
but cannot be eliminated.
}

\normalsize\baselineskip=15pt

\section{Introduction: Three-Neutrino Mixing Paradigm}

The results of several solar, atmospheric and long-baseline
neutrino oscillation experiment have proved that
neutrinos are massive and mixed particles
(see Ref.~\cite{Giunti-Kim-2007}).
There are two groups of experiments which measured
two independent squared-mass differences
($\Delta{m}^2$)
in two different neutrino flavor transition channels.

Solar neutrino experiments
(Homestake,
Kamiokande,
GALLEX/GNO,
SAGE,
Super-Kamiokande,
SNO,
BOREXino)
measured $\nu_{e} \to \nu_{\mu}, \nu_{\tau}$
oscillations generated by
$
\Delta m^2_{\text{SOL}}
=
6.2 {}^{+1.1}_{-1.9} \times 10^{-5} \, \text{eV}^2
$
and a mixing angle
$
\tan^2 \vartheta_{\text{SOL}}
=
0.42 {}^{+0.04}_{-0.02}
$
\cite{1010.0118}.
The KamLAND experiment
confirmed these oscillations by observing the disappearance
of reactor $\bar\nu_{e}$ at an average distance of about 180 km.
The combined fit of solar and KamLAND data leads to
$
\Delta m^2_{\text{SOL}}
=
(7.6 \pm 0.2) \times 10^{-5} \, \text{eV}^2
$
and a mixing angle
$
\tan^2 \vartheta_{\text{SOL}}
=
0.44 \pm 0.03
$
\cite{1010.0118}.
Notice that the agreement of solar and KamLAND data in favor of
$\nu_{e}$ and $\bar\nu_{e}$
disappearance generated by the same oscillation parameters
is consistent with the equality of
neutrino and antineutrino disappearance expected from
CPT symmetry
(see Ref.~\cite{Giunti-Kim-2007}).

Atmospheric neutrino experiments
(Kamiokande,
IMB,
Super-Kamiokande,
MACRO,
Soudan-2,
MINOS)
measured $\nu_{\mu}$ and $\bar\nu_{\mu}$
disappearance through oscillations generated by
$
\Delta m^2_{\text{ATM}}
\simeq
2.3 \times 10^{-3} \, \text{eV}^2
$
and a mixing angle
$
\sin^2 2\vartheta_{\text{ATM}}
\simeq
1
$
\cite{hep-ex/0501064}.
The K2K and MINOS long-baseline experiments
confirmed these oscillations by observing the disappearance
of accelerator $\nu_{\mu}$
at distances of about 250 km and 730 km, respectively.
The MINOS data give
$
\Delta m^2_{\text{ATM}}
=
2.32 {}^{+0.12}_{-0.08} \times 10^{-3} \, \text{eV}^2
$
and
$
\sin^2 2\vartheta_{\text{ATM}}
>
0.90
$
at 90\% C.L.
\cite{1103.0340}.
The equality of
muon neutrino and antineutrino disappearance expected from
CPT symmetry
is currently under investigation in the MINOS experiment
\cite{1104.0344},
with preliminary results which hint at an intriguing difference
between the muon neutrino and antineutrino
oscillation parameters.

These measurements led to the current
three-neutrino mixing paradigm,
in which the three active neutrinos
$\nu_{e}$,
$\nu_{\mu}$,
$\nu_{\tau}$
are superpositions of three massive neutrinos
$\nu_1$,
$\nu_2$,
$\nu_3$
with respective masses
$m_1$,
$m_2$,
$m_3$.
The two measured squared-mass differences can be interpreted as
\begin{equation}
\Delta m^2_{\text{SOL}}
=
\Delta m^2_{21}
\,,
\qquad
\Delta m^2_{\text{ATM}}
=
|\Delta m^2_{31}|
\simeq
|\Delta m^2_{32}|
\,,
\label{dm2}
\end{equation}
with
$\Delta m^2_{kj}=m_k^2-m_j^2$.
In the standard parameterization of the $3\times3$ unitary mixing matrix
(see Ref.~\cite{Giunti-Kim-2007})
$\vartheta_{\text{SOL}} \simeq \vartheta_{12}$,
$\vartheta_{\text{ATM}} \simeq \vartheta_{23}$
and
$\sin^2\vartheta_{13}<0.035$
at 90\% C.L.
\cite{0808.2016}.

\begin{figure}[t!]
\hfill
\begin{minipage}[l]{0.2\textwidth}
\begin{center}
\includegraphics*[bb=192 419 336 777, width=0.9\textwidth]{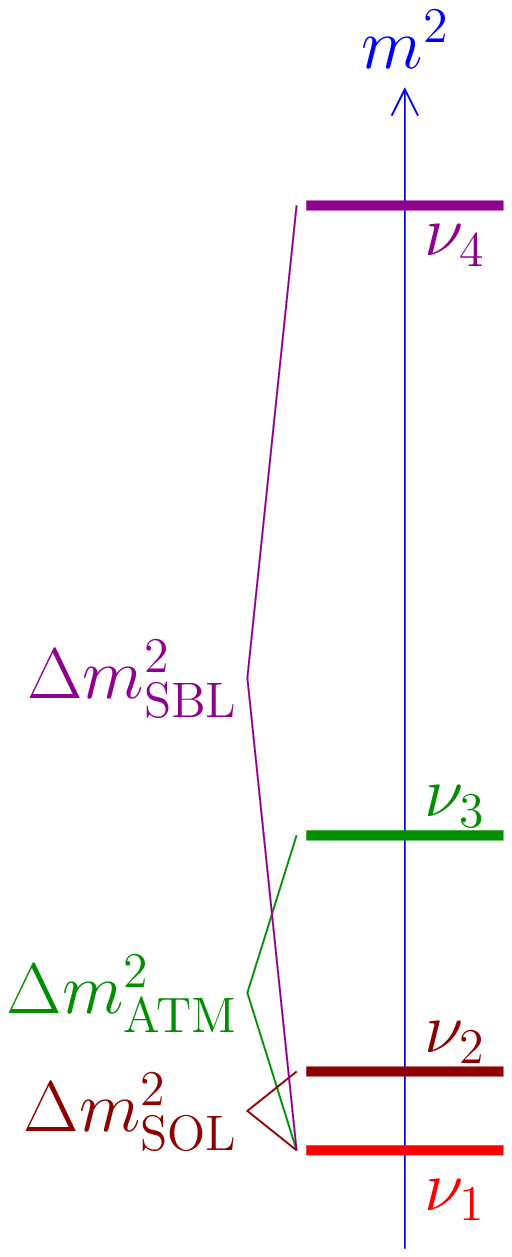}
\\
\text{{"normal"}}
\end{center}
\end{minipage}
\hfill
\begin{minipage}[l]{0.2\textwidth}
\begin{center}
\includegraphics*[bb=192 419 336 777, width=0.9\textwidth]{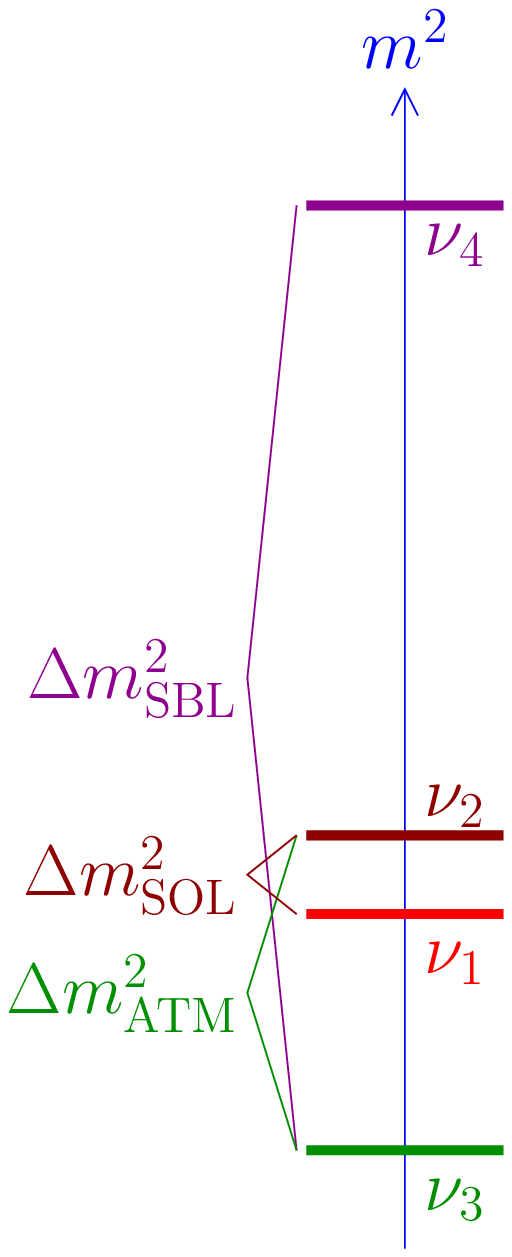}
\\
\text{{"3$\nu$-inverted"}}
\end{center}
\end{minipage}
\hfill
\begin{minipage}[l]{0.2\textwidth}
\begin{center}
\includegraphics*[bb=192 419 336 777, width=0.9\textwidth]{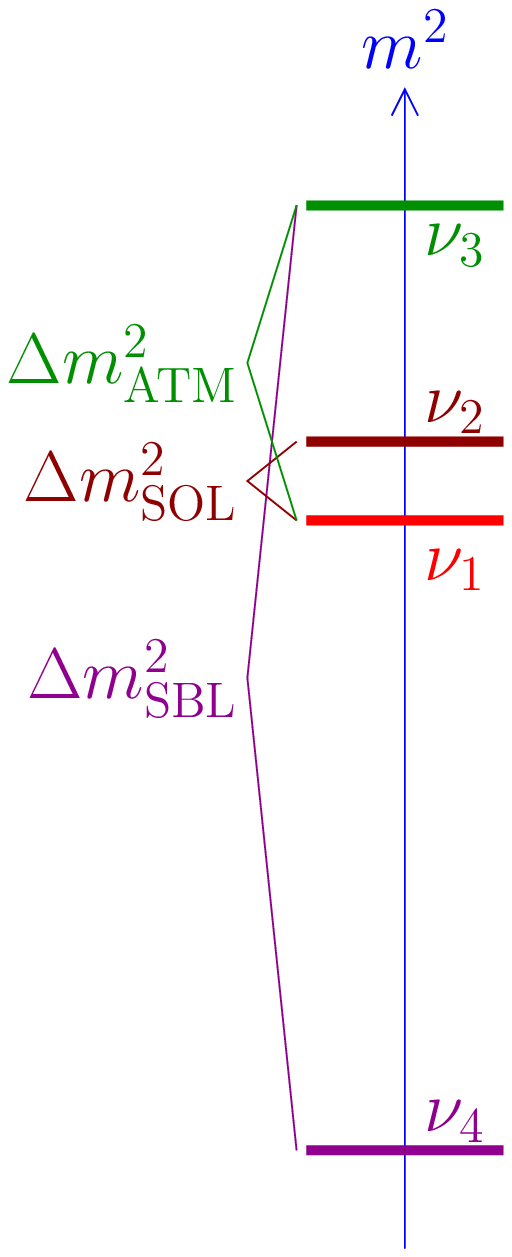}
\\
\text{{"4$\nu$-inverted"}}
\end{center}
\end{minipage}
\hfill
\begin{minipage}[l]{0.2\textwidth}
\begin{center}
\includegraphics*[bb=192 419 336 777, width=0.9\textwidth]{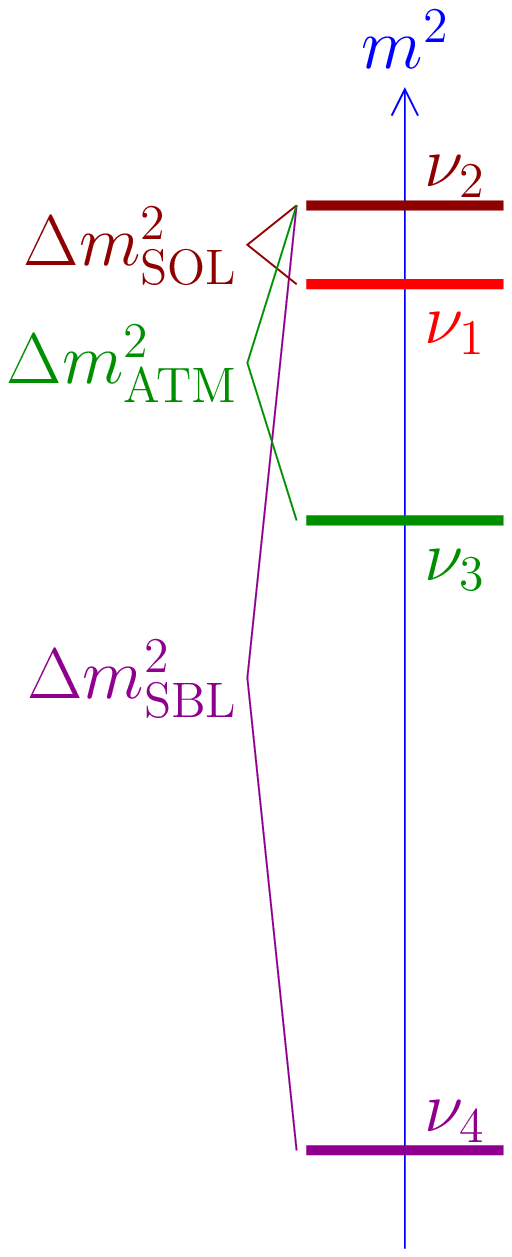}
\\
\text{{"totally-inverted"}}
\end{center}
\end{minipage}
\hfill
\caption{ \label{3+1}
3+1 four-neutrino schemes.
}
\end{figure}

The completeness of the three-neutrino mixing paradigm has been
challenged by the recent observation
of a signal of short-baseline
$\bar\nu_{\mu}\to\bar\nu_{e}$
oscillations in the MiniBooNE experiment
\cite{1007.1150}
which agrees with a similar signal
observed several years ago in the
LSND experiment
\cite{hep-ex/0104049}.
It is remarkable that the two signals have been observed at
different values of distance ($L$) and energy ($E$),
but approximately at the same $L/E$.
Since the distance and energy dependences of
neutrino oscillations occur through this ratio,
the agreement of the MiniBooNE and LSND signals raised
interest in the possibility of existence of
one or more squared-mass differences much larger than
$\Delta m^2_{\text{SOL}}$
and
$\Delta m^2_{\text{ATM}}$.
These new squared-mass differences
should have values larger than about 0.5 eV.

\section{3+1 Neutrino Mixing}

In the following, I consider first the
simplest extension of three-neutrino mixing
with the addition of one massive neutrino.
In such four-neutrino mixing framework
the flavor neutrino basis is composed by the three active neutrinos
$\nu_{e}$,
$\nu_{\mu}$,
$\nu_{\tau}$
and
a sterile neutrino
$\nu_{s}$
which does not have weak interactions and does not contribute
to the invisible width of the $Z$ boson
\cite{hep-ex/0509008}.
The existence of sterile neutrinos
which have been thermalized in the early Universe is compatible
with Big-Bang Nucleosynthesis data
\cite{astro-ph/0408033,1001.4440}
and cosmological measurements of the
Cosmic Microwave Background and Large-Scale Structures
if the mass of the fourth neutrino is limited below about 1 eV
\cite{1006.5276,1102.4774}.

\begin{figure}[t!]
\begin{center}
\begin{tabular}{cc}
Old reactor $\bar\nu_{e}$ fluxes
&
New reactor $\bar\nu_{e}$ fluxes
\\
\includegraphics*[bb=15 14 573 562, width=0.45\textwidth]{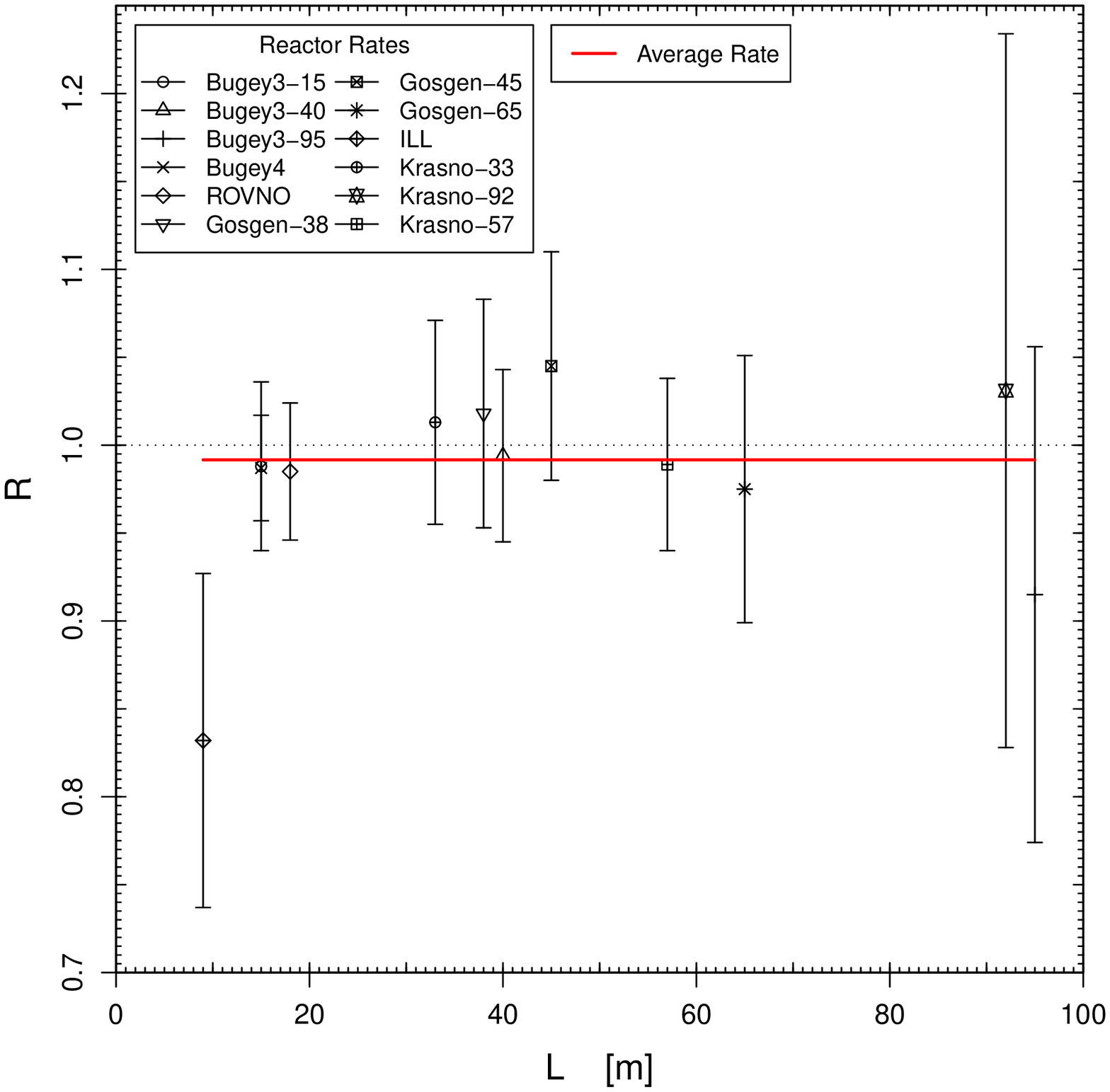}
&
\includegraphics*[bb=15 14 573 562, width=0.45\textwidth]{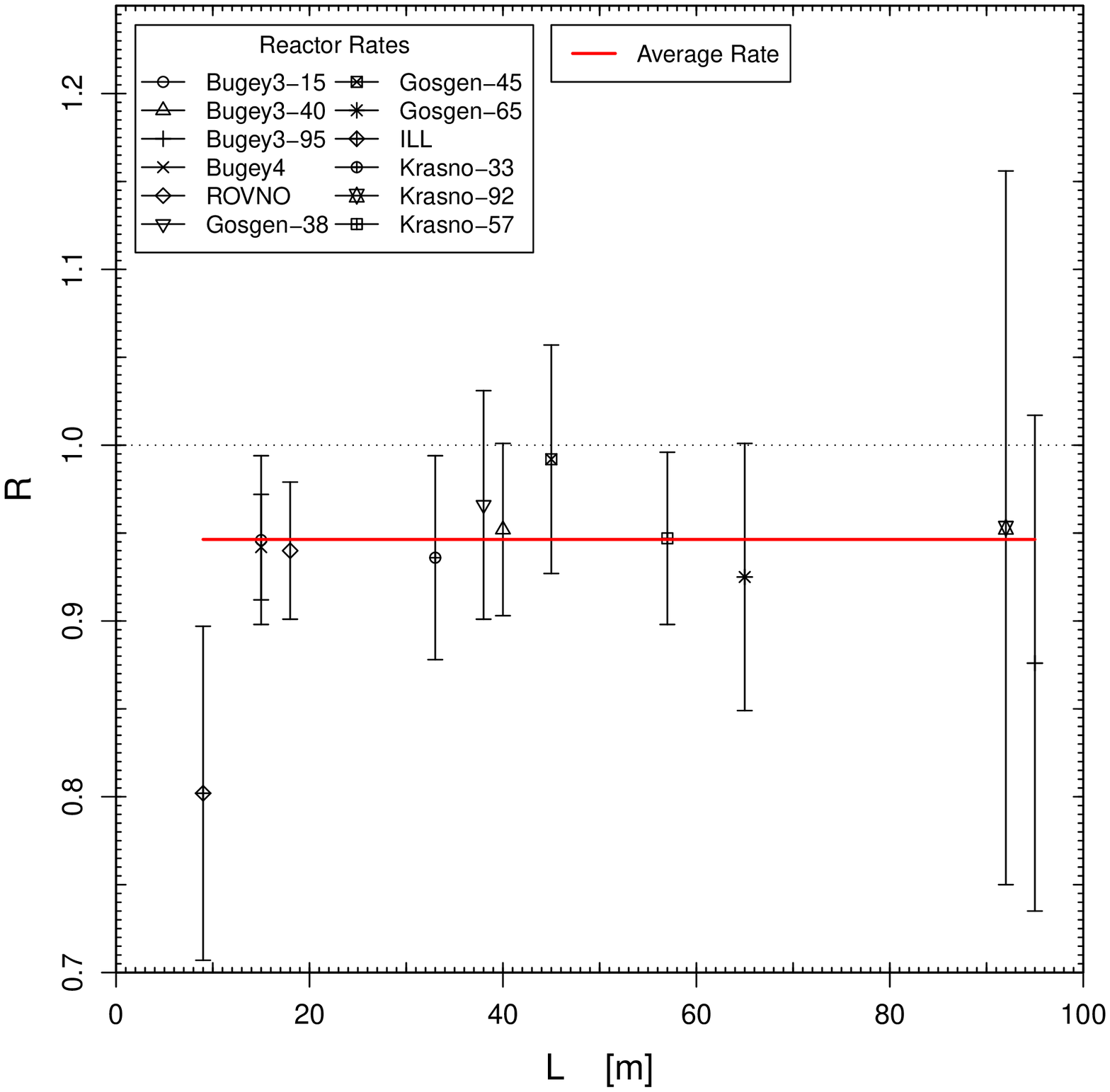}
\\[-1.8cm]
\fbox{\ensuremath{\displaystyle
\overline{R}
=
0.992 \pm 0.024
}}
&
\fbox{\ensuremath{\displaystyle
\overline{R}
=
0.946 \pm 0.024
}}
\end{tabular}
\end{center}
\vspace{1cm}
\caption{ \label{reactors}
Ratio $R$ of the observed $\bar\nu_{e}$ event rate and that expected in absence of $\bar\nu_{e}$ disappearance
obtained from the
old
(see Ref.~\cite{hep-ph/0107277})
and
new
\cite{1101.2663}
reactor $\bar\nu_{e}$ fluxes.
The average value of $R$ obtained with the
new reactor $\bar\nu_{e}$ fluxes quantifies the
reactor antineutrino anomaly \cite{1101.2755}.
}
\end{figure}

So-called 2+2 four-neutrino mixing schemes are strongly disfavored
by the absence of any signal of sterile neutrino effects in
solar and atmospheric neutrino data
\cite{hep-ph/0405172}.
Hence, we must consider the so-called 3+1 four-neutrino schemes
depicted in Fig.~\ref{3+1}.
Since the
"4$\nu$-inverted"
and
"totally-inverted"
schemes have three massive neutrinos at the eV scale,
they are disfavored by cosmological data
over the
"normal"
and
"3$\nu$-inverted"
schemes.
In all 3+1 schemes the
effective flavor transition and survival probabilities
in short-baseline (SBL) experiments
are given by
\begin{align}
\null & \null
P_{\boss{\nu}{\alpha}\to\boss{\nu}{\beta}}^{\text{SBL}}
=
\sin^{2} 2\vartheta_{\alpha\beta}
\sin^{2}\left( \frac{\Delta{m}^{2} L}{4E} \right)
\qquad
(\alpha\neq\beta)
\,,
\label{trans}
\\
\null & \null
P_{\nu_{\alpha}\to\nu_{\alpha}}^{\text{SBL}}
=
1
-
\sin^{2} 2\vartheta_{\alpha\alpha}
\sin^{2}\left( \frac{\Delta{m}^{2} L}{4E} \right)
\,,
\label{survi}
\end{align}
for
$\alpha,\beta=e,\mu,\tau,s$,
with
$\Delta{m}^{2} = \Delta{m}^{2}_{\text{SBL}}$
and
\begin{align}
\null & \null
\sin^{2} 2\vartheta_{\alpha\beta}
=
4 |U_{\alpha4}|^2 |U_{\beta4}|^2
\,,
\label{transsin}
\\
\null & \null
\sin^{2} 2\vartheta_{\alpha\alpha}
=
4 |U_{\alpha4}|^2 \left( 1 - |U_{\alpha4}|^2 \right)
\,.
\label{survisin}
\end{align}
Therefore:
\begin{enumerate}
\item
All effective SBL oscillation probabilities
depend only on the largest squared-mass difference
$\Delta{m}^{2} = \Delta{m}^{2}_{\text{SBL}} = |\Delta{m}^{2}_{41}|$.
\item
All oscillation channels are open, each one with its own oscillation amplitude.
\item
All oscillation amplitudes depend only on the absolute values
of the elements in the fourth column of the mixing matrix,
i.e. on three real numbers with sum less than unity,
since the unitarity of the mixing matrix implies
$
\sum_{\alpha} |U_{\alpha4}|^2 = 1
$
\item
CP violation cannot be observed in SBL oscillation experiments,
even if the mixing matrix contains CP-violation phases.
In other words,
neutrinos and antineutrinos have the same
effective SBL oscillation probabilities.
\end{enumerate}

\begin{figure}[t!]
\begin{center}
\begin{tabular}{cc}
\includegraphics*[bb=7 14 563 569, width=0.45\textwidth]{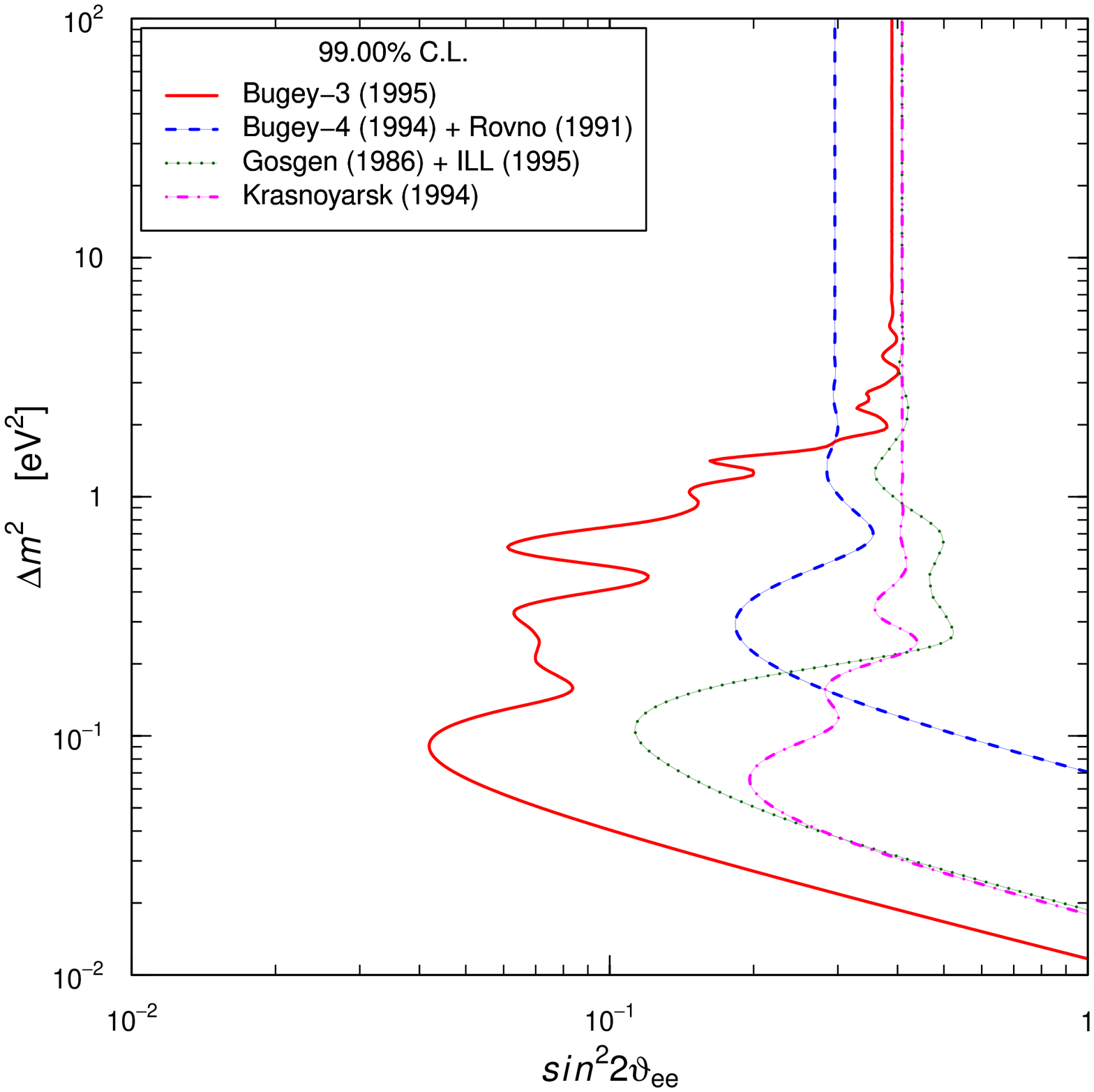}
&
\includegraphics*[bb=7 14 563 569, width=0.45\textwidth]{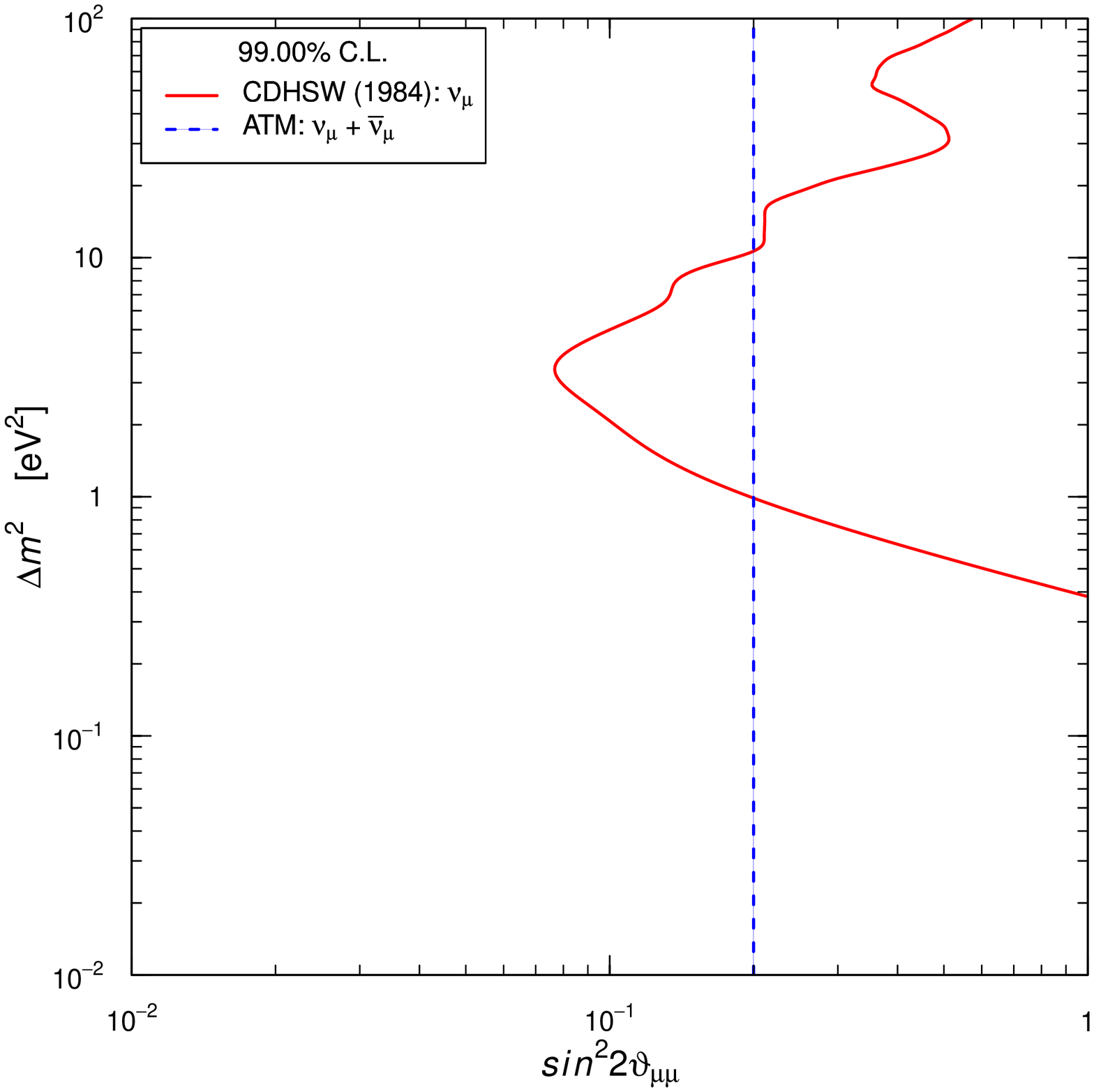}
\end{tabular}
\end{center}
\caption{ \label{dis}
Exclusion curves obtained from the data of reactor $\bar\nu_{e}$ disappearance experiments
(see Ref.~\cite{1101.2755}),
from the data of the CDHSW $\nu_{\mu}$ disappearance experiment
\cite{Dydak:1984zq},
and from atmospheric neutrino data
(extracted from the analysis in Ref.~\cite{0705.0107}).
}
\end{figure}

Before the recent indication of an antineutrino
$\bar\nu_{\mu}\to\bar\nu_{e}$
signal consistent with the
LSND antineutrino signal,
the MiniBooNE collaboration published
the results of neutrino data which do not show
a corresponding $\nu_{\mu}\to\nu_{e}$
signal
\cite{0812.2243}.
This difference between the MiniBooNE neutrino and antineutrino data may be due to CP violation.

The absence of any difference in the
effective SBL oscillation probabilities
of
neutrinos and antineutrinos
in 3+1 four-neutrino mixing schemes
implies that these schemes
cannot explain the difference between
neutrinos and antineutrino oscillations observed in the
MiniBooNE.
Moreover,
the dependence of all the oscillation amplitudes in Eqs.~(\ref{transsin}) and (\ref{survisin})
on three independent absolute values
of the elements in the fourth column of the mixing matrix
implies that the amplitude of
$
\boss{\nu}{\mu}\to\boss{\nu}{e}
$
transitions is limited by the absence of large SBL disappearance of
$\boss{\nu}{e}$ and $\boss{\nu}{\mu}$
observed in several experiments.

The results of reactor neutrino experiments constrain the value
$|U_{e4}|^2$
through the measurement of
$\sin^{2} 2\vartheta_{ee}$.
The calculation of the reactor $\bar\nu_{e}$ flux has been recently improved in Ref.~\cite{1101.2663},
resulting in an increase of about 3\% with respect to the previous value adopted
by all experiments for the comparison with the data.
As illustrated in Fig.~\ref{reactors},
the measured reactor rates are in agreement with those derived from the old $\bar\nu_{e}$ flux,
but show a deficit of about $2.2\sigma$ with respect to the rates derived from the new $\bar\nu_{e}$ flux.
This is the ``reactor antineutrino anomaly'' \cite{1101.2755}\footnote{
We do not consider here the ``Gallium neutrino anomaly''
\cite{hep-ph/0610352,0707.4593,0711.4222,0902.1992,1005.4599,1006.3244},
which may be compatible with the reactor antineutrino anomaly
assuming the equality of neutrino and antineutrino disappearance imposed by the CPT symmetry.
},
which may be an indication in the
$\bar\nu_{e}\to\bar\nu_{e}$
channel of a signal corresponding to the
$\bar\nu_{\mu}\to\bar\nu_{e}$
signal observed in the
LSND and MiniBooNE experiments.
However,
the $\bar\nu_{e}$ disappearance is small and large values of
$\sin^{2} 2\vartheta_{ee}$
are constrained by the exclusion curves in the left panel of Fig.~\ref{dis}.
Since values of $|U_{e4}|^2$ close to unity are excluded by solar neutrino oscillations
(which require large $|U_{e1}|^2+|U_{e2}|^2$),
for small $\sin^{2} 2\vartheta_{ee}$ we have
\begin{equation}
\sin^{2} 2\vartheta_{ee} \simeq 4 |U_{e4}|^2
\,.
\label{ue4}
\end{equation}

The value of $\sin^{2} 2\vartheta_{\mu\mu}$
is constrained by the curves in the right panel of Fig.~\ref{dis},
which have been obtained from
the lack of $\nu_{\mu}$ disappearance in the CDHSW $\nu_{\mu}$ experiment
\cite{Dydak:1984zq}
and
from the requirement of large $|U_{\mu1}|^2+|U_{\mu2}|^2+|U_{\mu3}|^2$
for atmospheric neutrino oscillations \cite{0705.0107}.
Hence,
$|U_{\mu4}|^2$ is small and
\begin{equation}
\sin^{2} 2\vartheta_{\mu\mu} \simeq 4 |U_{\mu4}|^2
\,.
\label{um4}
\end{equation}

\begin{figure}[t!]
\begin{center}
\begin{tabular}{cc}
\includegraphics*[bb=7 14 563 569, width=0.45\textwidth]{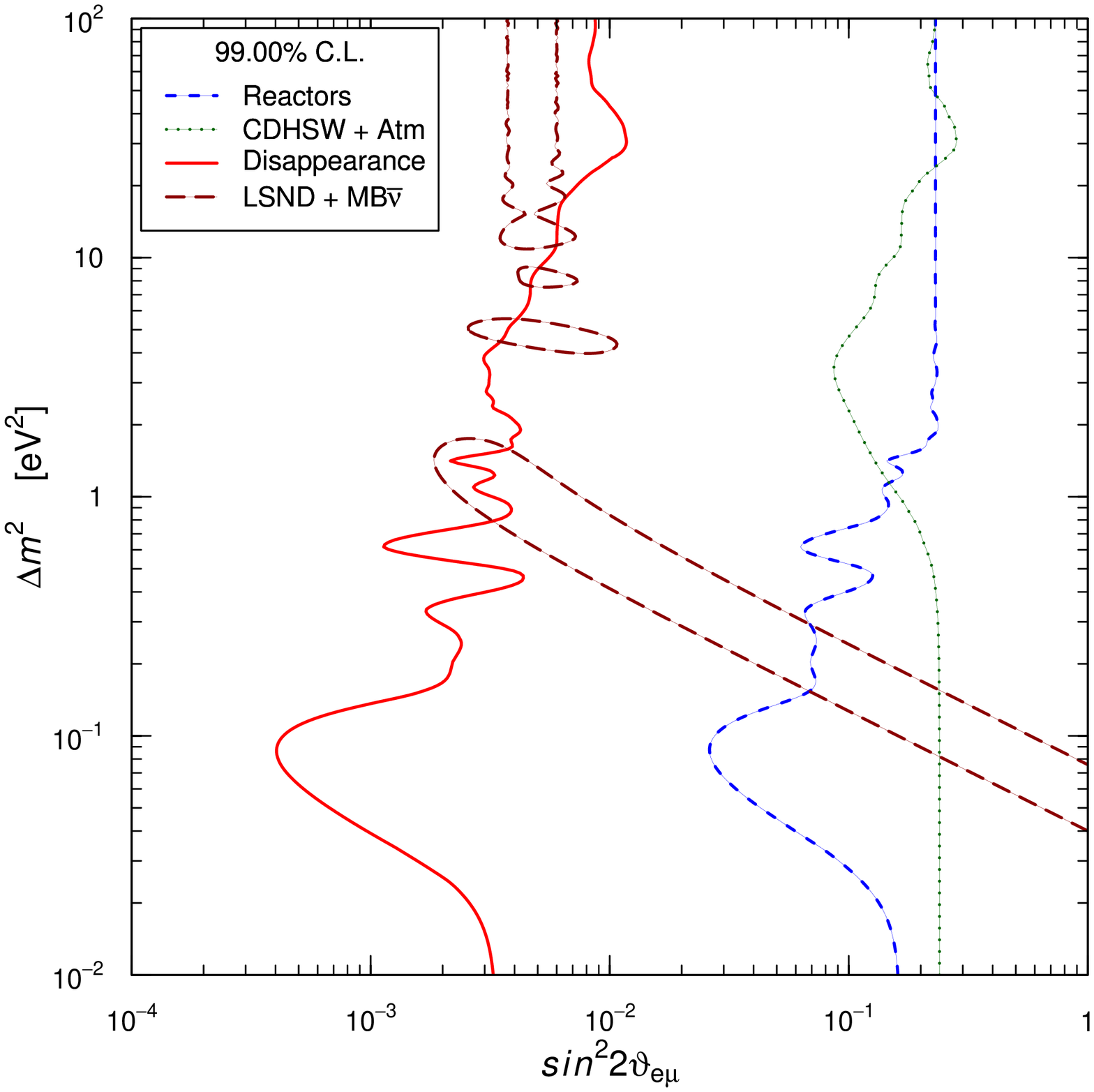}
&
\includegraphics*[bb=7 14 563 569, width=0.45\textwidth]{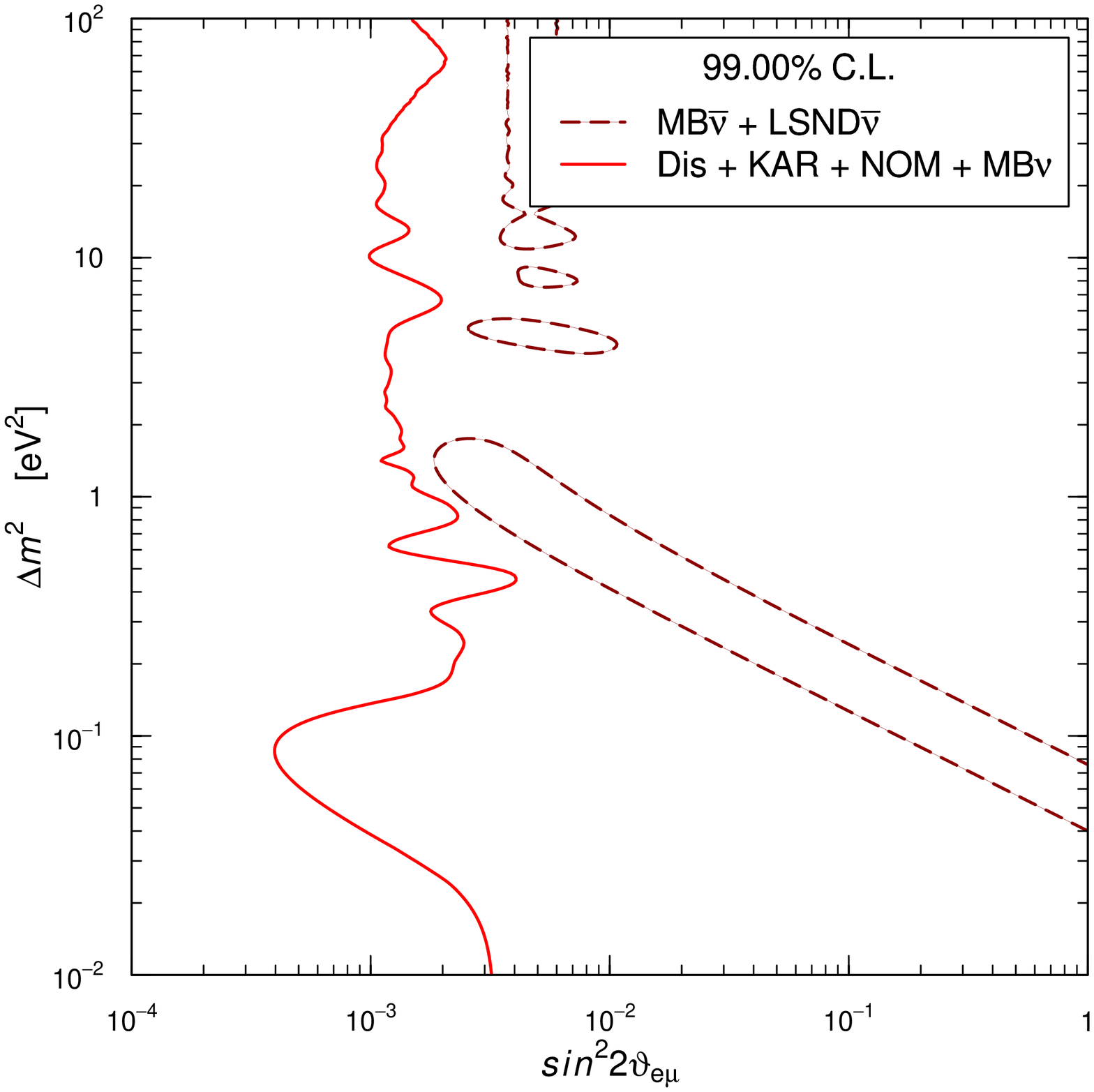}
\end{tabular}
\end{center}
\caption{ \label{exc}
Left Panel:
Exclusion curves
in the $\sin^{2} 2\vartheta_{e\mu}$--$\Delta{m}^2$ plane
obtained from the separate constraints in Fig.~\ref{dis}
(blue and green lines)
and the combined constraint given by Eq.~(\ref{sem})
(red line)
from disappearance experiments (Dis).
Right Panel:
Exclusion curve obtained with the addition of
KARMEN \cite{hep-ex/0203021} (KAR),
NOMAD \cite{hep-ex/0306037} (NOM)
and 
MiniBooNE neutrino \cite{0812.2243} (MB$\nu$)
data (red line).
In both panels the region enclosed by the dark-red lines
is allowed by
LSND and MiniBooNE antineutrino data.
}
\end{figure}

From Eqs.~(\ref{transsin}), (\ref{ue4}) and (\ref{um4}),
for the amplitude of
$
\boss{\nu}{\mu}\to\boss{\nu}{e}
$
transitions we obtain
\begin{equation}
\sin^{2} 2\vartheta_{e\mu}
\simeq
\frac{1}{4}
\,
\sin^{2} 2\vartheta_{ee}
\,
\sin^{2} 2\vartheta_{\mu\mu}
\,.
\label{sem}
\end{equation}
Therefore,
if
$\sin^{2} 2\vartheta_{ee}$
and
$\sin^{2} 2\vartheta_{\mu\mu}$
are small,
$\sin^{2} 2\vartheta_{e\mu}$
is quadratically suppressed.
This is illustrated in the left panel of Fig.~\ref{exc},
where one can see that the separate effects of the constraints on
$\sin^{2} 2\vartheta_{ee}$
and
$\sin^{2} 2\vartheta_{\mu\mu}$
exclude only the large-$\sin^{2} 2\vartheta_{e\mu}$
part of the region allowed by
LSND and MiniBooNE antineutrino data,
whereas most of this region is excluded by the combined constraint in Eq.~(\ref{sem}).
As shown in the right panel of Fig.~\ref{exc},
the constraint becomes stronger by including the data of the
KARMEN \cite{hep-ex/0203021},
NOMAD \cite{hep-ex/0306037}
and 
MiniBooNE neutrino \cite{0812.2243}
experiments,
which did not observe a short-baseline
$
\boss{\nu}{\mu}\to\boss{\nu}{e}
$
signal.
Since the parameter goodness-of-fit
\cite{hep-ph/0304176}
is
0.0016\%,
3+1 schemes are strongly disfavored by the data.
This conclusion has been reached recently also in Refs.~\cite{0705.0107,1007.4171,1012.0267,1103.4570}
and confirms the pre-MiniBooNE results in Refs.~\cite{hep-ph/0207157,hep-ph/0405172}.

\begin{figure}[t!]
\begin{center}
\begin{tabular}{ccc}
\includegraphics*[bb=14 14 563 549, width=0.3\textwidth]{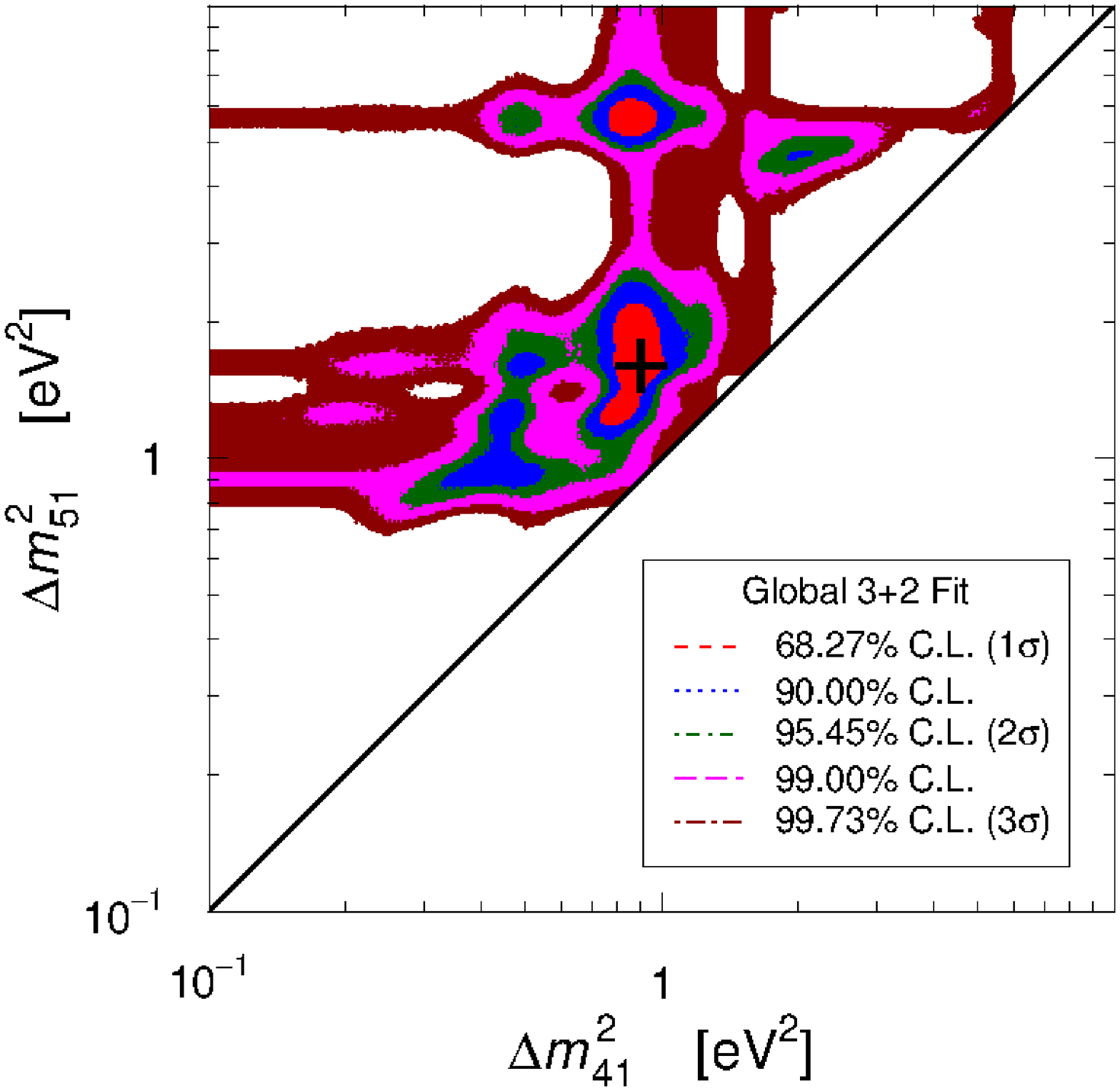}
&
\includegraphics*[bb=14 14 563 549, width=0.3\textwidth]{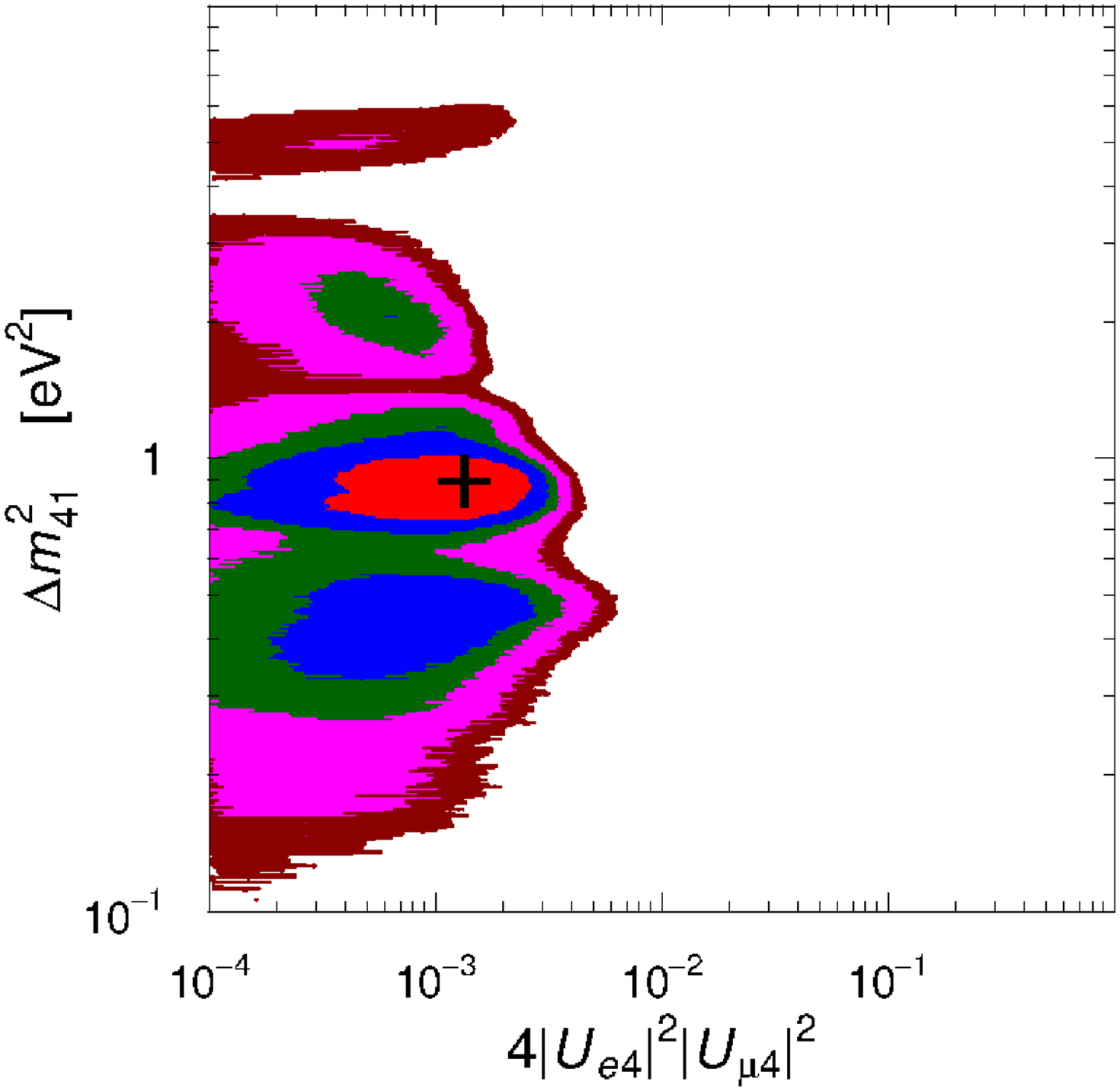}
&
\includegraphics*[bb=14 14 563 549, width=0.3\textwidth]{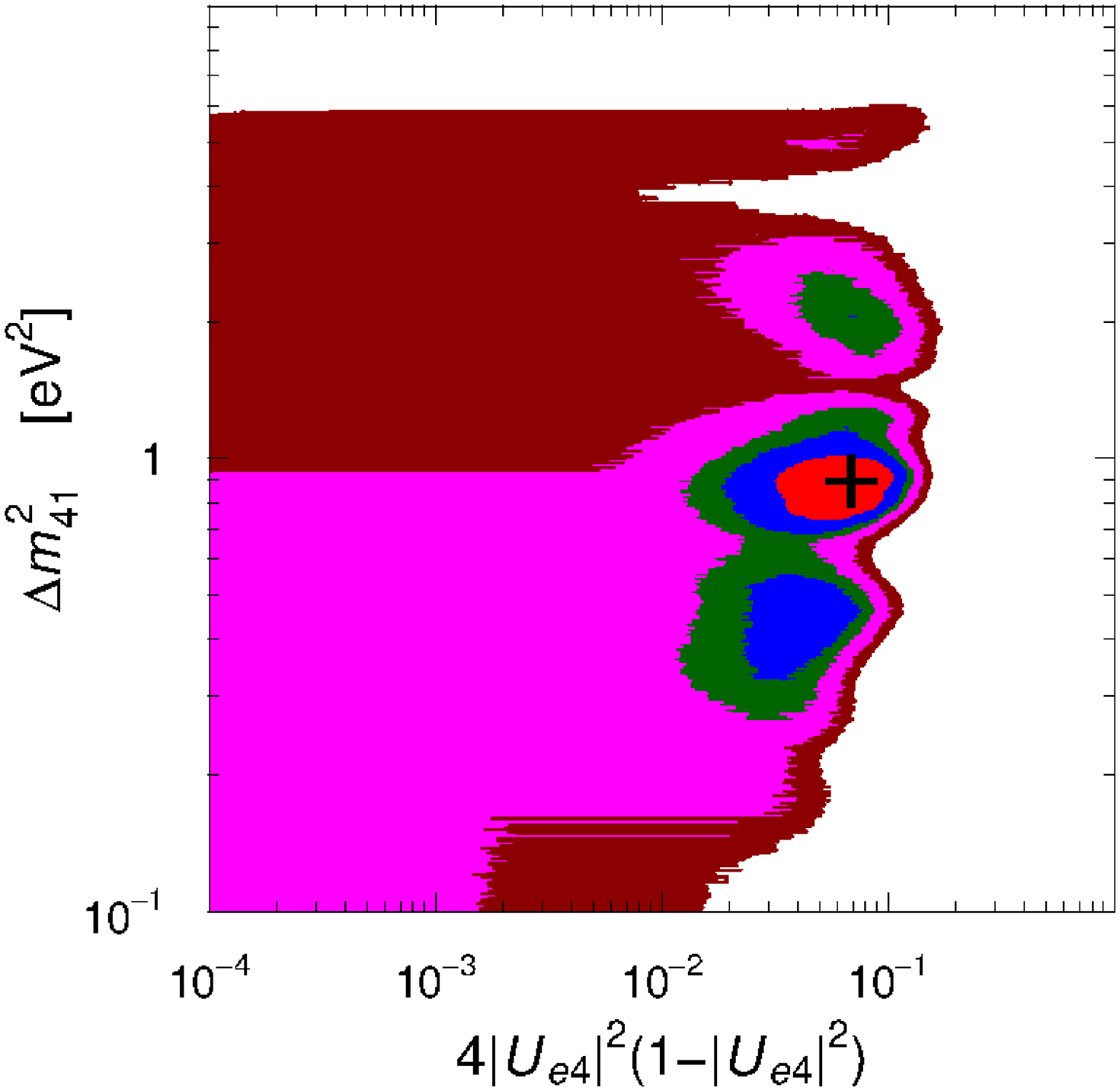}
\\
\includegraphics*[bb=14 14 563 549, width=0.3\textwidth]{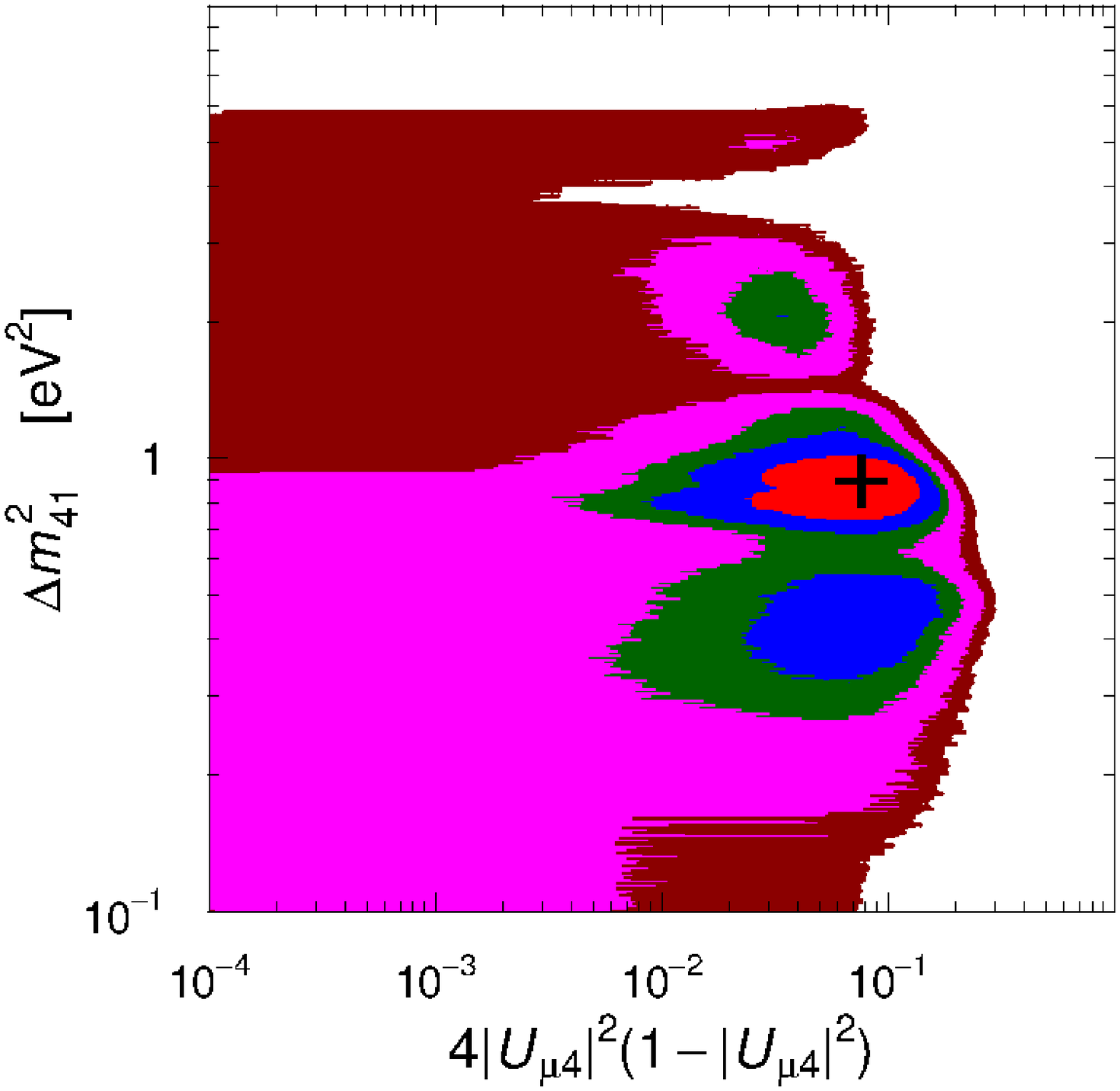}
&
\includegraphics*[bb=14 14 563 549, width=0.3\textwidth]{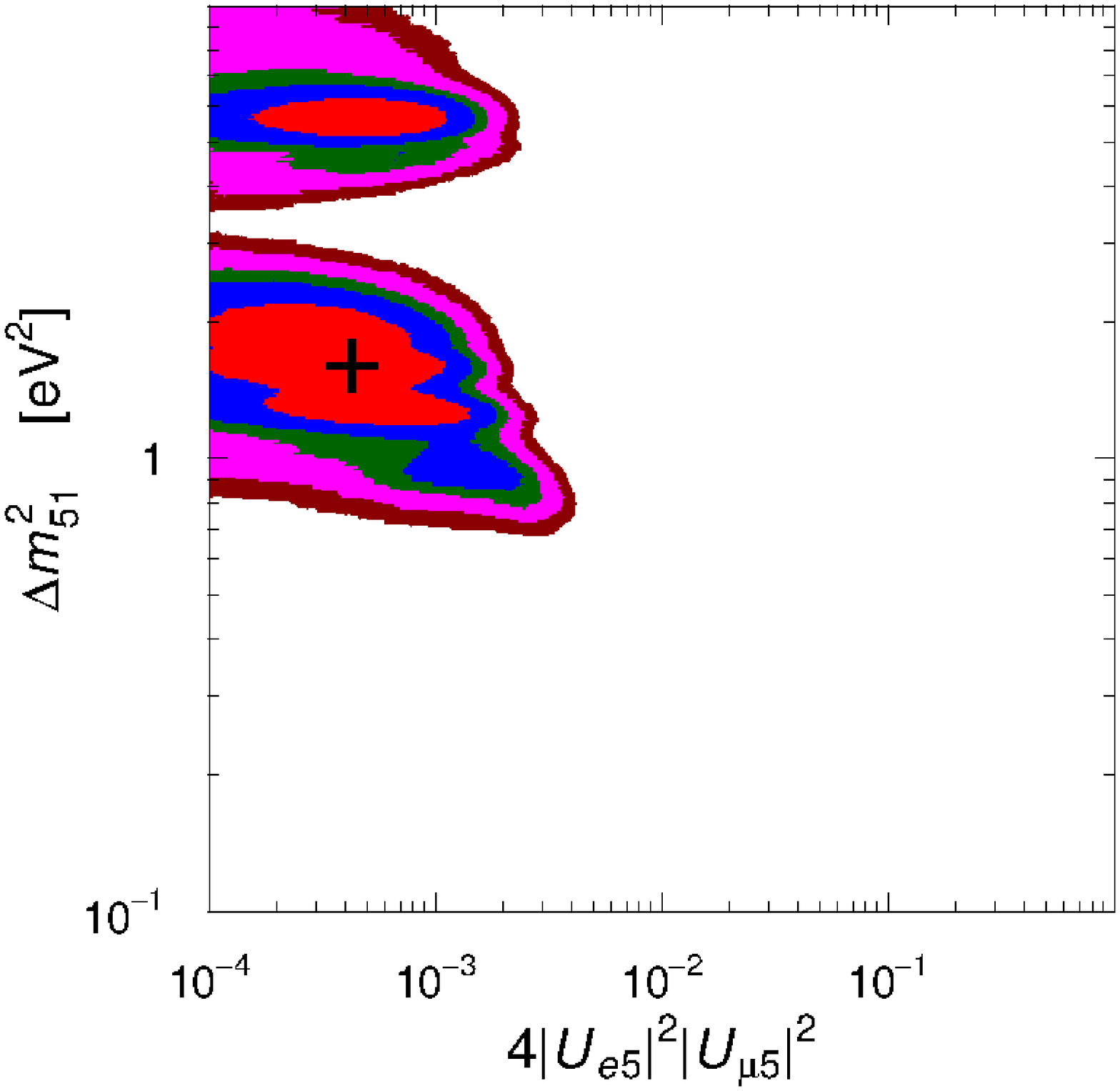}
&
\includegraphics*[bb=14 14 563 549, width=0.3\textwidth]{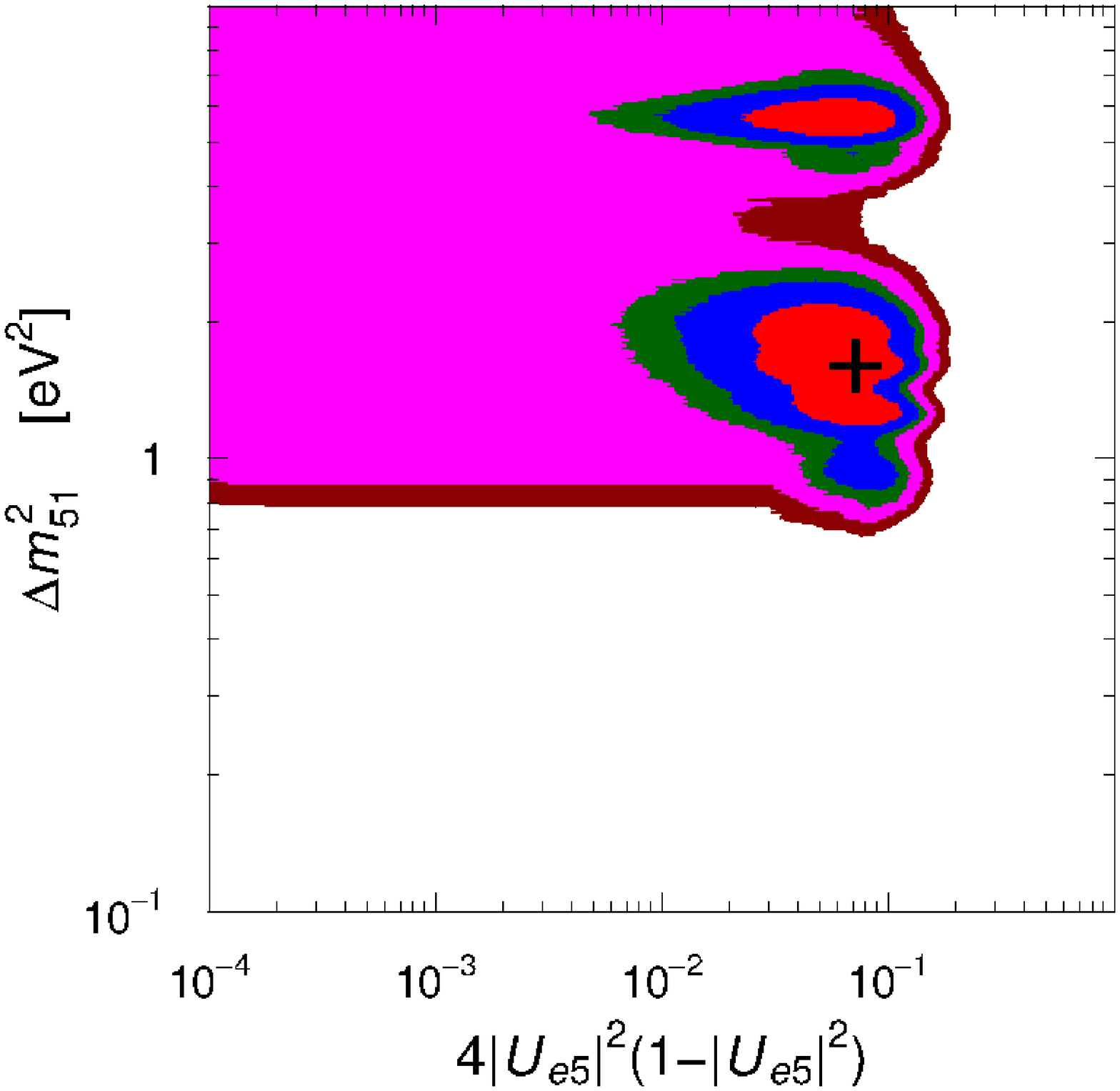}
\\
\includegraphics*[bb=14 14 563 549, width=0.3\textwidth]{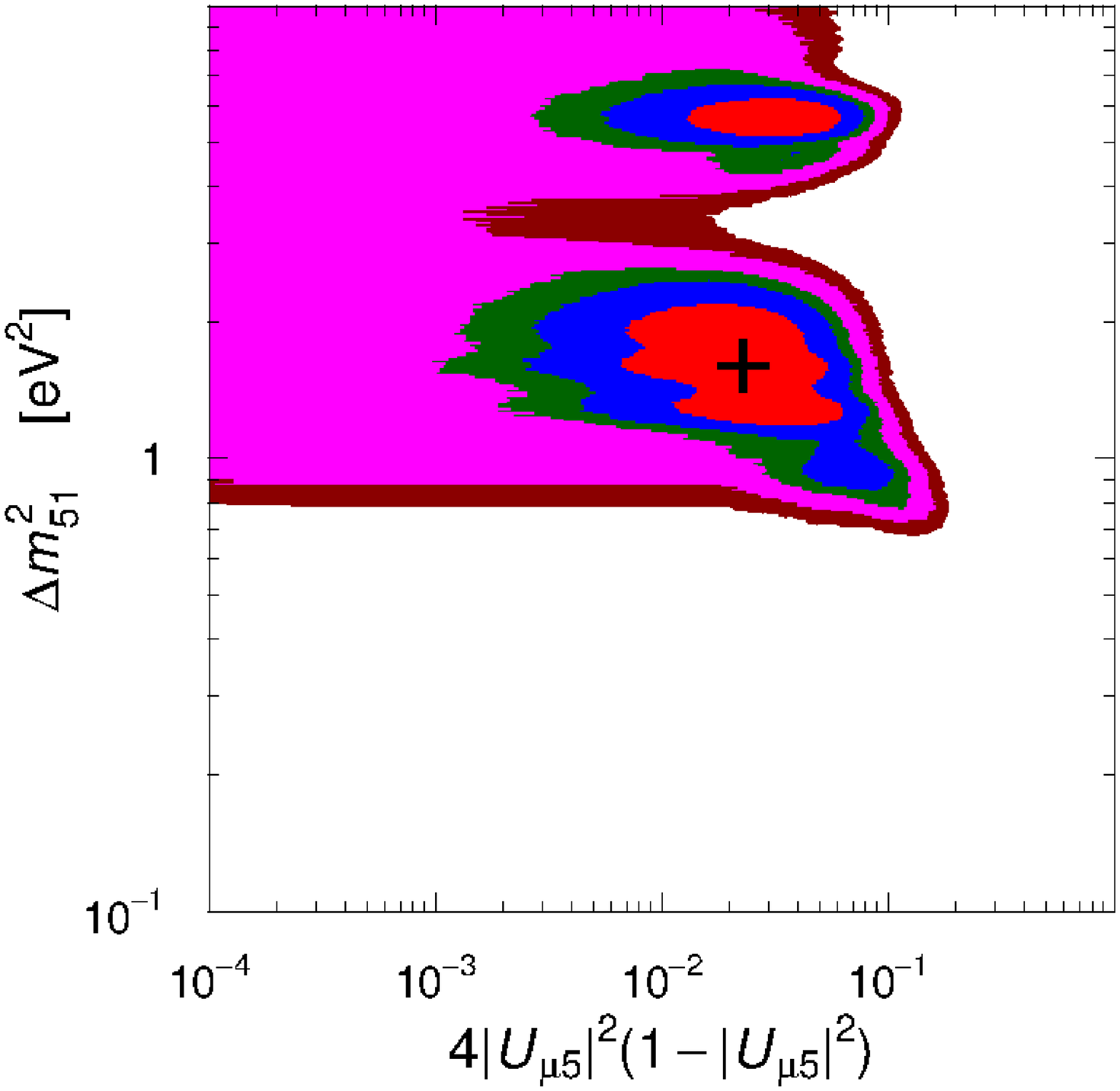}
&
\includegraphics*[bb=14 14 563 549, width=0.3\textwidth]{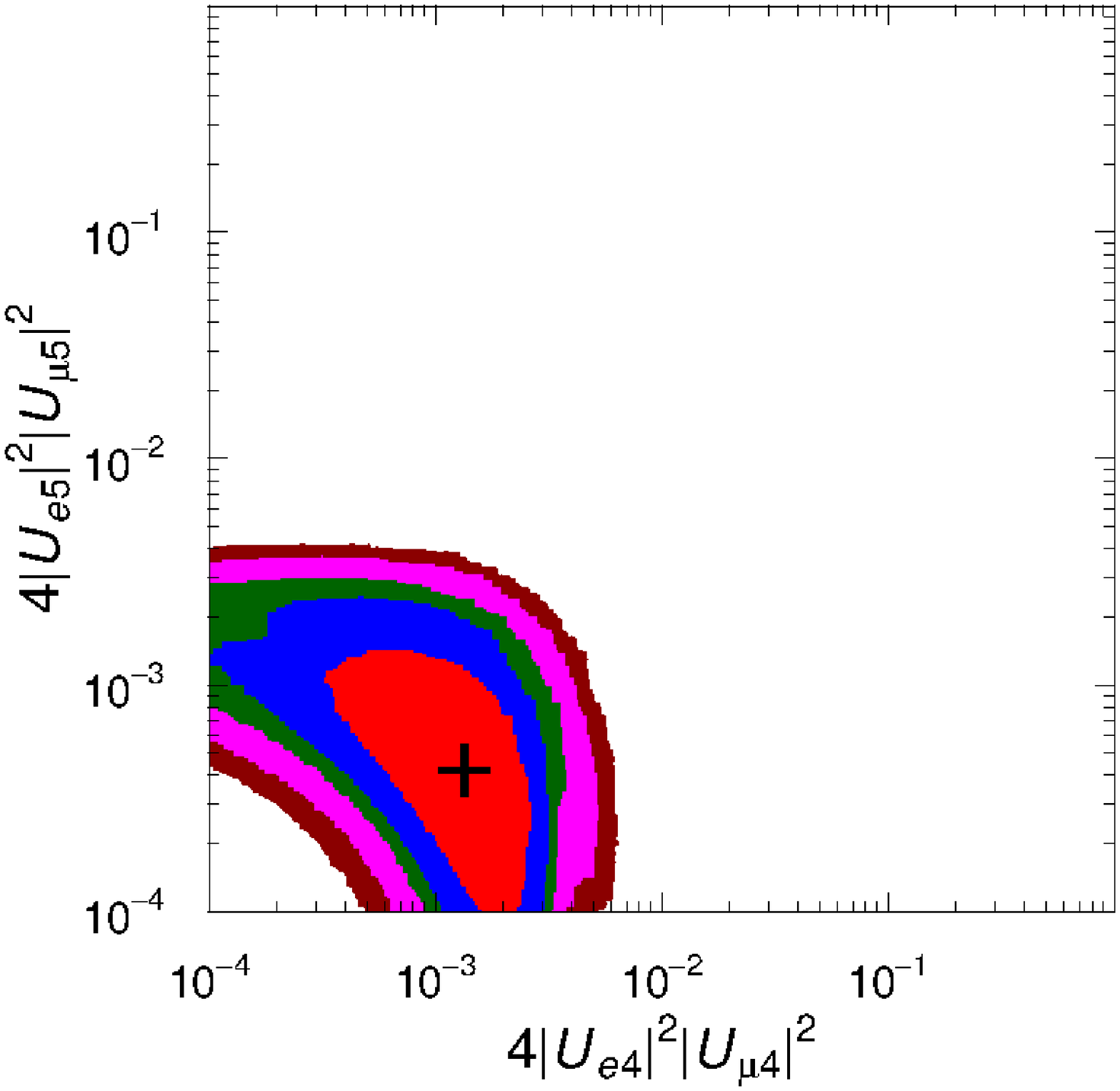}
&
\includegraphics*[bb=14 14 563 549, width=0.3\textwidth]{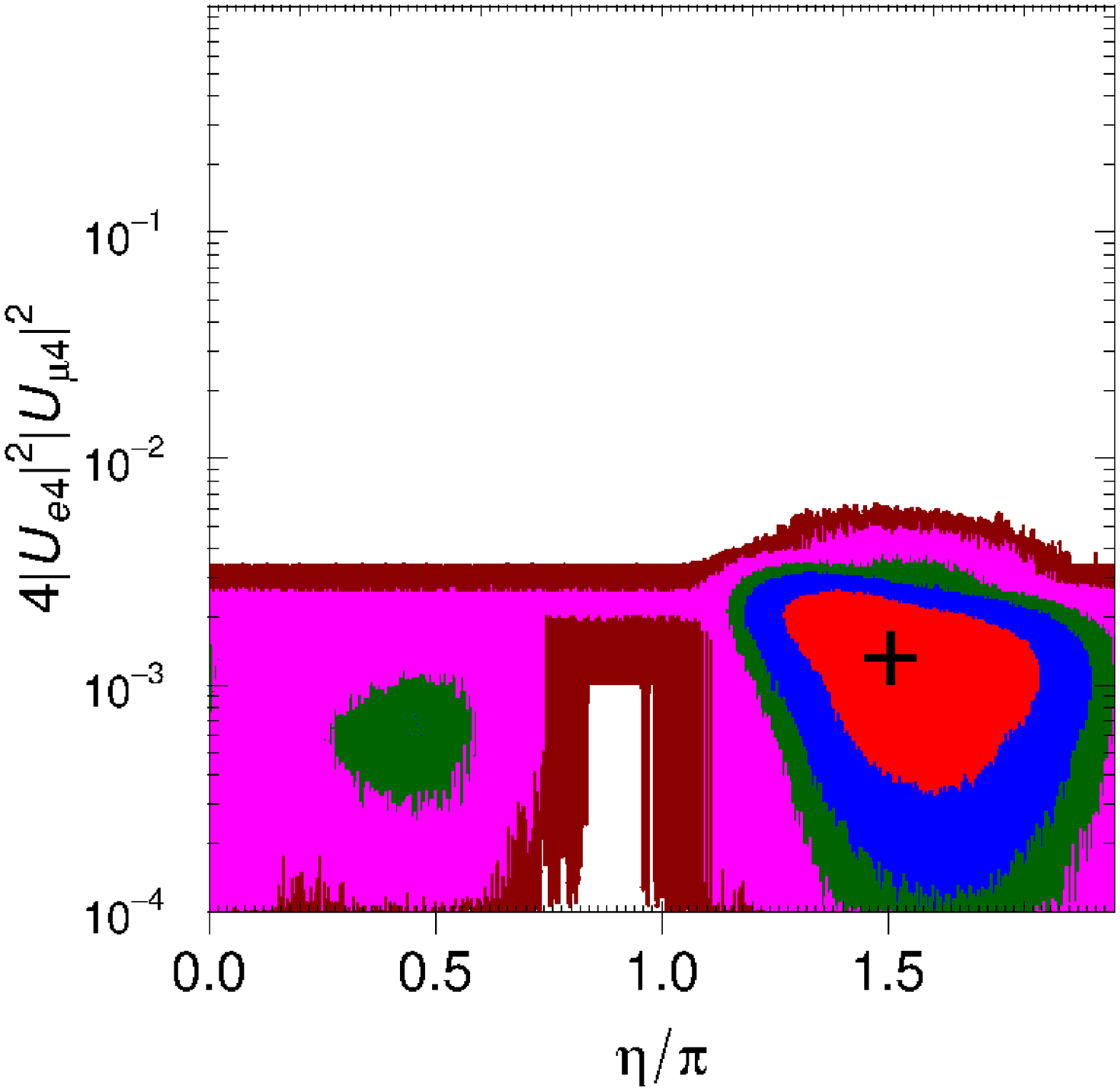}
\end{tabular}
\end{center}
\caption{ \label{fig-3p2}
Marginal allowed regions in two-dimensional planes of interesting combinations of the
oscillation parameters in 3+2 neutrino mixing.
}
\end{figure}

The CP-violating difference between MiniBooNE neutrino and antineutrino data
can be explained by introducing another physical effect in addition to a sterile neutrino:
a second sterile neutrino in 3+2 schemes \cite{hep-ph/0305255,hep-ph/0609177,0906.1997,0705.0107,1007.4171,1103.4570},
non-standard interactions \cite{1007.4171},
CPT violation \cite{1010.1395,1012.0267}.
In the following I discuss the possibility of 3+2 neutrino mixing.

\section{3+2 Neutrino Mixing}

In 3+2 schemes the relevant effective oscillation probabilities in short-baseline experiments are given by
\begin{align}
P_{\boss{\nu}{\mu}\to\boss{\nu}{e}}^{\text{SBL}}
=
\null & \null
4 |U_{\mu4}|^2 |U_{e4}|^2 \sin^{2}\phi_{41}
+
4 |U_{\mu5}|^2 |U_{e5}|^2 \sin^{2}\phi_{51}
\\
\null & \null
+
8 |U_{\mu4} U_{e4} U_{\mu5} U_{e5}|
\sin\phi_{41}
\sin\phi_{51}
\cos(\phi_{54} \stackrel{(+)}{-} \eta)
\,,
\nonumber
\label{trans-3p2}
\\
P_{\nu_{\alpha}\to\nu_{\alpha}}^{\text{SBL}}
=
\null & \null
1
-
4 (1 - |U_{\alpha4}|^2 - |U_{\alpha5}|^2)
(|U_{\alpha4}|^2 \sin^{2}\phi_{41} + |U_{\alpha5}|^2 \sin^{2}\phi_{51})
\\
\null & \null
+ 4 |U_{\alpha4}|^2 |U_{\alpha5}|^2 \sin^{2}\phi_{54}
\,,
\nonumber
\label{survi-3p2}
\end{align}
for
$\alpha,\beta=e,\mu$,
with
\begin{equation}
\phi_{kj}
=
\Delta{m}^2_{kj} L / 4 E
\,,
\qquad
\eta
=
\text{arg}[U_{e4}^{*}U_{\mu4}U_{e5}U_{\mu5}^{*}]
\,.
\label{3p2}
\end{equation}
Note the change in sign of the contribution of the CP-violating phase $\eta$
going from neutrinos to antineutrinos,
which allows us
to explain the CP-violating difference between MiniBooNE neutrino and antineutrino data.

Figure~\ref{fig-3p2}
shows the marginal allowed regions in two-dimensional planes of interesting combinations of the
oscillation parameters in our 3+2 global fit of the same set of data used in Fig.~\ref{exc}.
The best-fit values of the mixing parameters are:
\begin{eqnarray}
&&
\Delta{m}^2_{41} = 0.90 \, \text{eV}^2
\,,\,
|U_{e4}|^2 = 0.017
\,,\,
|U_{\mu4}|^2 = 0.019
\,,
\label{bef1}
\\
&&
\Delta{m}^2_{51} = 1.61 \, \text{eV}^2
\,,\,
|U_{e5}|^2 = 0.018
\,,\,
|U_{\mu5}|^2 = 0.0058
\,,\,
\eta = 1.51 \pi
\,.
\label{bef2}
\end{eqnarray}
The parameter goodness-of-fit obtained with the comparison of the fit of
LSND and MiniBooNE antineutrino data
and the fit of all other data is
0.24\%.
This is an improvement with respect to the
0.0016\%
parameter goodness-of-fit obtained in 3+1 schemes.
However,
the value of the parameter goodness-of-fit
remains low as a consequence of the
fact that the
$\bar\nu_{\mu}\to\bar\nu_{e}$
transitions observed in LSND and MiniBooNE
must correspond in any neutrino mixing schemes to
enough short-baseline disappearance of $\boss{\nu}{e}$ and $\boss{\nu}{\mu}$
which has not been observed.

The results of our 3+2 global fit are in reasonable agreement with those presented in
Ref.~\cite{1103.4570}.
There is a discrepancy in the location of the best-fit point in the
$\Delta{m}^2_{41}$--$\Delta{m}^2_{51}$
plane,
but we obtain similar regions for the local $\chi^2$ minima.
Our allowed regions are larger than those presented in
Ref.~\cite{1103.4570}.
I think that such difference is probably due to a different treatment of the
spectral data of the Bugey-3 reactor experiment \cite{Declais:1995su}
which cause the wiggling
for
$\Delta{m}^2 \lesssim 1 \, \text{eV}^2$
of the disappearance limit
in the left panel of Fig.~\ref{exc}
and the exclusion curve
in the right panel of Fig.~\ref{exc}.
Such wiggling is wider in Fig.~3 of Ref.~\cite{1103.4570},
leading to deeper valleys of the $\chi^2$ function
and
smaller allowed regions.

\section{Conclusions}

In conclusion,
I think that
we are living an exciting time in neutrino physics
which may prelude to a transition from the well-established
three-neutrino mixing paradigm to a new paradigm
of neutrino mixing with sterile neutrinos
and possibly other effects (as non-standard interactions and CPT violation)
which are very interesting for the exploration of the physics beyond the Standard Model.
In order to clarify the validity of the experimental indications in favor of an expansion
of neutrino mixing beyond the standard three-neutrino mixing
and resolve the tension between the current positive and negative experimental results,
new experiments with high sensitivity and low background are needed
(see, for example, those proposed in
Refs.~\cite{hep-ex/9901012,0909.0355,1006.2103,1007.3228,1011.4509,1103.5307}).

\bibliography{%
bibtex/nu%
}

\begin{thebibliography}{10}

\bibitem{Giunti-Kim-2007}
C. Giunti and C.W. Kim,
{Fundamentals of Neutrino Physics and Astrophysics} (Oxford
University Press, Oxford, UK, 2007).

\bibitem{1010.0118}
Super-Kamiokande, K. Abe et~al.,
Phys. Rev. D83 (2011) 052010, arXiv:1010.0118.

\bibitem{hep-ex/0501064}
Super-Kamiokande, Y. Ashie et~al.,
Phys. Rev. D71 (2005) 112005, hep-ex/0501064.

\bibitem{1103.0340}
MINOS, P. Adamson et~al.,
Phys. Rev. Lett. 106 (2011) 181801, arXiv:1103.0340.

\bibitem{1104.0344}
MINOS, P. Adamson et~al.,
(2011), arXiv:1104.0344.

\bibitem{0808.2016}
T. Schwetz, M. Tortola and J.W.F. Valle,
New J. Phys. 10 (2008) 113011, arXiv:0808.2016.

\bibitem{1007.1150}
MiniBooNE, A.A. Aguilar-Arevalo et~al.,
Phys. Rev. Lett. 105 (2010) 181801, arXiv:1007.1150.

\bibitem{hep-ex/0104049}
LSND, A. Aguilar et~al.,
Phys. Rev. D64 (2001) 112007, hep-ex/0104049.

\bibitem{hep-ex/0509008}
ALEPH, DELPHI, L3, OPAL, SLD, LEP Electroweak Working Group, SLD Electroweak
Group, SLD Heavy Flavour Group, S. Schael et~al.,
Phys. Rept. 427 (2006) 257, hep-ex/0509008.

\bibitem{astro-ph/0408033}
R.H. Cyburt et~al.,
Astropart. Phys. 23 (2005) 313, astro-ph/0408033.

\bibitem{1001.4440}
Y.I. Izotov and T.X. Thuan,
Astrophys. J. 710 (2010) L67, arXiv:1001.4440.

\bibitem{1006.5276}
J. Hamann et~al.,
Phys. Rev. Lett. 105 (2010) 181301, arXiv:1006.5276.

\bibitem{1102.4774}
E. Giusarma et~al.,
(2011), arXiv:1102.4774.

\bibitem{hep-ph/0107277}
C. Bemporad, G. Gratta and P. Vogel,
Rev. Mod. Phys. 74 (2002) 297, hep-ph/0107277.

\bibitem{1101.2663}
T.A. Mueller et~al.,
(2011), arXiv:1101.2663.

\bibitem{1101.2755}
G. Mention et~al.,
Phys. Rev. D83 (2011) 073006, arXiv:1101.2755.

\bibitem{hep-ph/0405172}
M. Maltoni et~al.,
New J. Phys. 6 (2004) 122, hep-ph/0405172.

\bibitem{Dydak:1984zq}
CDHSW, F. Dydak et~al.,
Phys. Lett. B134 (1984) 281.

\bibitem{0705.0107}
M. Maltoni and T. Schwetz,
Phys. Rev. D76 (2007) 093005, arXiv:0705.0107.

\bibitem{0812.2243}
MiniBooNE, A.A. Aguilar-Arevalo,
Phys. Rev. Lett. 102 (2009) 101802, arXiv:0812.2243.

\bibitem{hep-ph/0610352}
C. Giunti and M. Laveder,
Mod. Phys. Lett. A22 (2007) 2499, hep-ph/0610352.

\bibitem{0707.4593}
C. Giunti and M. Laveder,
Phys. Rev. D77 (2008) 093002, arXiv:0707.4593.

\bibitem{0711.4222}
M.A. Acero, C. Giunti and M. Laveder,
Phys. Rev. D78 (2008) 073009, arXiv:0711.4222.

\bibitem{0902.1992}
C. Giunti and M. Laveder,
Phys. Rev. D80 (2009) 013005, arXiv:0902.1992.

\bibitem{1005.4599}
C. Giunti and M. Laveder,
Phys. Rev. D82 (2010) 053005, arXiv:1005.4599.

\bibitem{1006.3244}
C. Giunti and M. Laveder,
(2010), arXiv:1006.3244.

\bibitem{hep-ex/0203021}
KARMEN, B. Armbruster et~al.,
Phys. Rev. D65 (2002) 112001, hep-ex/0203021.

\bibitem{hep-ex/0306037}
NOMAD, P. Astier et~al.,
Phys. Lett. B570 (2003) 19, hep-ex/0306037.

\bibitem{hep-ph/0304176}
M. Maltoni and T. Schwetz,
Phys. Rev. D68 (2003) 033020, hep-ph/0304176.

\bibitem{1007.4171}
E. Akhmedov and T. Schwetz,
JHEP 10 (2010) 115, arXiv:1007.4171.

\bibitem{1012.0267}
C. Giunti and M. Laveder,
Phys. Rev. D83 (2011) 053006, arXiv:1012.0267.

\bibitem{1103.4570}
J. Kopp, M. Maltoni and T. Schwetz,
(2011), arXiv:1103.4570.

\bibitem{hep-ph/0207157}
M. Maltoni et~al.,
Nucl. Phys. B643 (2002) 321, hep-ph/0207157.

\bibitem{hep-ph/0305255}
M. Sorel, J. Conrad and M. Shaevitz,
Phys. Rev. D70 (2004) 073004, hep-ph/0305255.

\bibitem{hep-ph/0609177}
G. Karagiorgi et~al.,
Phys. Rev. D75 (2007) 013011, hep-ph/0609177.

\bibitem{0906.1997}
G. Karagiorgi et~al.,
Phys. Rev. D80 (2009) 073001, arXiv:0906.1997.

\bibitem{1010.1395}
C. Giunti and M. Laveder,
Phys. Rev. D82 (2010) 093016, arXiv:1010.1395.

\bibitem{Declais:1995su}
Bugey, B. Achkar et~al.,
Nucl. Phys. B434 (1995) 503.

\bibitem{hep-ex/9901012}
A. Ianni, D. Montanino and G. Scioscia,
Eur. Phys. J. C8 (1999) 609, hep-ex/9901012.

\bibitem{0909.0355}
B. Baibussinov et~al.,
(2009), arXiv:0909.0355.

\bibitem{1006.2103}
V.N. Gavrin et~al.,
(2010), arXiv:1006.2103.

\bibitem{1007.3228}
S.K. Agarwalla and P. Huber,
Phys. Lett. B696 (2011) 359, arXiv:1007.3228.

\bibitem{1011.4509}
S.K. Agarwalla and R.S. Raghavan,
(2010), arXiv:1011.4509.

\bibitem{1103.5307}
J. Vergados, Y. Giomataris and Y. Novikov,
(2011), arXiv:1103.5307.

\end{thebibliography}

\end{document}